\documentclass[12pt]{article}
\usepackage{amsmath,amssymb,amsfonts}
\usepackage{graphicx}
\usepackage{hyperref}
\hypersetup{colorlinks,linkcolor={blue},citecolor={blue},urlcolor={red}} 
\usepackage[english]{babel}
\usepackage{bm}
\usepackage[utf8]{inputenc}
\usepackage{listings}
\usepackage[T1]{fontenc}
\usepackage{lmodern}
\usepackage{caption}
\usepackage{multirow}
\usepackage{makecell}
\usepackage[normalem]{ulem}
\usepackage{floatrow}
\usepackage{subcaption}
\usepackage{mathrsfs}
\usepackage{enumerate}
\usepackage{color}
\usepackage{array}
\usepackage{float}
\usepackage{feynmp}
\usepackage{tikz-feynman}
\tikzfeynmanset{compat=1.1.0}
\usepackage{csquotes}

\usepackage{xcolor}
\usepackage[table]{xcolor}
\usepackage{pifont}
\usepackage[
backend=biber,
style=numeric-comp,
sorting=none
]{biblatex}
\begin{document}

\title{Neutrino Masses and Dark Matter Stability in 3HDMs with Minimal Non-Abelian Discrete Symmetries}

\date{}
\author{
\\[1mm] 
Cesar Bonilla$^{}$\footnote{E-mail: {\tt cesar.bonilla@ucn.cl }}~
and Andr\'es~Layana-Ram\'irez$^{}$\footnote{E-mail: {\tt andres.layana@alumnos.ucn.cl}}
\\[3mm]
	\textit{\small Departamento de F\'isica, Universidad Cat\'olica del Norte,}\\
	\textit{\small Avenida Angamos 0610, Antofagasta, 1240000, Chile}\\[3mm]
    }
\maketitle

\begin{abstract}
\noindent
We present minimal three-Higgs-doublet models (3HDMs) based on global non-Abelian
discrete symmetries that simultaneously explain neutrino masses and dark matter
stability. A residual $Z_2$ parity from the spontaneous breaking of the new
symmetry stabilizes the dark matter candidate, which runs in the loop generating
neutrino masses at one loop alongside a tree-level type-I seesaw contribution.
We identify $S_3$ and $D_4$ as the smallest non-Abelian groups realizing this
framework and determine the minimal models in both cases that are consistent with
current neutrino oscillation data. The resulting models conserve CP in both the
Yukawa and scalar sectors.
\end{abstract}

\maketitle

\section{Introduction}
\noindent

Two of the most compelling pieces of evidence for physics beyond the Standard
Model (SM) are the observed oscillations of neutrinos between flavor states and
the existence of non-baryonic dark matter (DM). Neutrino oscillations require at
least two non-zero neutrino masses~\cite{deSalas:2020pgw,Esteban:2024eli}, while dark matter
requires an electrically neutral particle that is stable on cosmic timescales and absent from the SM spectrum. Addressing both shortcomings in a unified and economical framework is a central
goal of model building beyond the SM~\cite{Avila:2025qsc}.
\\

The simplest mechanism for generating small neutrino masses is the type-I seesaw,
which introduces heavy right-handed (RH) neutrinos~\cite{Minkowski:1977sc, GellMann:1979vob,Yanagida:1979as,Mohapatra:1979ia, Schechter:1980gr}.
A single RH neutrino $N_1$ produces a rank-1 mass matrix, accounting for only
one non-zero mass eigenvalue and therefore only for the atmospheric mass scale
$\Delta m^2_{\rm atm}$. In order to explain the other mass scale, one could introduce an additional RH neutrino \cite{Schechter:1980gr}, as this would allow the generation of two non-vanishing mass eigenvalues. To explain the full oscillation picture, including the
distinct solar scale $\Delta m^2_{\rm sol} \ll \Delta m^2_{\rm atm}$, the minimal
setup must be extended. 
One natural and predictive way to generate both scales is to attribute them to distinct mass mechanisms. That is, the atmospheric scale arises at tree level via a type-I seesaw, while the solar scale is generated radiatively through corrections mediated by particles belonging to a stable dark sector running in the loop, whose lightest component constitutes the dark matter candidate~\cite{Tao:1996vb,Ma:2006km,Rojas:2018wym,Aranda:2018lif}\footnote{For a recent 
comprehensive review on the connection between neutrino mass generation via a dark sector see~\cite{Avila:2025qsc}}. The latter 
can be either of fermionic or bosonic nature, depending on the mass spectrum of the 
dark sector. In this picture, the dark matter candidate is naturally embedded in the 
mechanism responsible for generating the solar neutrino mass scale.
\\

A deeper question is the origin of the stabilizing symmetry. Rather than imposing
a $Z_2$ by hand, it is more appealing to derive it as the residual subgroup of a
spontaneously broken global symmetry that independently accounts for the observed
pattern of neutrino mixing angles. This idea known as the {\it discrete dark matter}
(DDM) mechanism was first proposed in~\cite{Hirsch:2010ru} and further
developed in~\cite{Boucenna:2011tj}, where the flavor group $A_4$ spontaneously
breaks to a $Z_2$ parity that simultaneously stabilizes the dark matter and
constrains the neutrino sector. While the original realization of~\cite{Hirsch:2010ru}
predicted a vanishing reactor mixing angle $\theta_{13}$ and is therefore disfavored
by current oscillation data, the underlying mechanism remains attractive and has
motivated subsequent extensions. In~\cite{Bonilla:2023pna} this
mechanism was combined with the type-I seesaw and one-loop radiative mass
generation within an $A_4$ model, requiring a four-Higgs-doublet structure (4HDM)
and three RH neutrinos since $A_4$ is the smallest non-Abelian group with a
three-dimensional irreducible representation.
\\

Following this reasoning a natural question arises, {\it what is the minimal group structure that can
realize the DDM mechanism with type-I seesaw and one-loop radiative mass generation?} Groups smaller than $A_4$ whose largest
irreducible representation is two-dimensional necessarily lead to three-Higgs-doublet
models (3HDMs). From these, one is SM-like singlet $H_1$ and a pair $\Phi=(H_2,\eta)^T$
transforming as the two-dimensional representation. The two smallest such groups
are $S_3$ (order 6) and $D_4$ (order 8). Moving from $A_4$ to these groups reduces
the scalar sector from four to three doublets, a non-trivial step since the DDM
mechanism imposes specific conditions on the vacuum expectation value alignment and residual symmetry.
\\

In this work we propose 3HDMs based on the smallest 
non-Abelian discrete symmetries admitting the DDM mechanism, $G_D \in \{S_3, D_4\}$. 
We find that reproducing all neutrino oscillation observables requires extending the 
naively minimal field content associated with $G_D$ by one further representation, 
$N_S$, in addition to the scalar $G_D$-doublet $\Phi$ and the Majorana $G_D$-doublet 
$N_D$. In all viable realizations CP is conserved in both the Yukawa and scalar 
sectors.
\\

The paper is organized as follows. Section~\ref{sec:min} outlines the general
requirements and imposed constraints for a minimal DDM 3HDM. Sections~\ref{sec:s3} and~\ref{sec:d4}
present the $S_3$ and $D_4$ realizations, including the numerical results. We conclude in Section~\ref{sec:conc}.

\section{Minimal Models}
\label{sec:min}
\noindent

Let us first outline the requirements for constructing a minimal non-Abelian
discrete dark matter model. In this context, minimality entails the use of a
single non-Abelian symmetry group $G_D$ smaller than
$A_4$~\cite{Hirsch:2010ru,Bonilla:2023pna}\footnote{Throughout this 
work, \emph{minimal} refers to the choice of the smallest non-Abelian group $G_D$ 
realizing the DDM mechanism, not to the particle content.}, focusing on groups whose largest
irreducible representation is two-dimensional~\cite{Ishimori:2010au}. The scalar
sector must include an $SU(2)$ doublet, $H_1 \simeq \mathbf{1} \in G_D$, with
trivial charge under $G_D$ to generate the SM fermion masses.
\\

The stability of the dark sector arises from the breaking of $G_D$ through additional fields
transforming non-trivially under the flavor symmetry. The non-trivial scalar is
$\Phi = (H_2, \eta)^T \simeq \mathbf{2} \in G_D$, where both $H_2$ and $\eta$
are $SU(2)$ doublets. The model, therefore, necessarily contains three $SU(2)$
doublets. That is, one singlet $H_1$ and one flavor-doublet $\Phi$. This leads to the 3HDM
structure. To prevent flavor-changing neutral currents, $\Phi$ does not couple
to quarks.
\\

The non-Abelian group $G_D$ breaks partially when $\Phi$ acquires a vacuum expectation value (vev), leaving
a non-trivial residual subgroup $G_R \subset G_D$:
\begin{equation}
    G_D \xrightarrow{\,\,\langle \Phi \rangle\,\,} G_R,
\end{equation}
where $G_R$ stabilizes the DM candidate\footnote{The minimal discrete symmetry
capable of stabilizing a dark matter candidate is a $Z_2$ parity}. We choose
the vev alignment $\langle \Phi \rangle = (\langle H_2 \rangle, 0)^T$, that is, $\eta$ is a vevless field and carries a non-trivial
$G_R$ charge. This vacuum alignment is invariant
by a specific generator of $G_D$. 
\\

\begin{figure}[h!]
    \centering
   \includegraphics[width=0.85\textwidth]{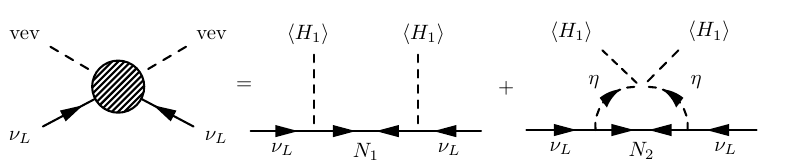}
    \vspace{5mm}
    \caption{\small Neutrino mass contributions at tree and loop level
    as in~\cite{Rojas:2018wym,Aranda:2018lif}.}
    \label{o5seesaws}
\end{figure}

As mentioned, at least two RH neutrinos $(N_1, N_2)$ are required to account for both neutrino mass scales~\cite{Rojas:2018wym,Aranda:2018lif}. We take this pair of Majorana fermions to form a doublet of the new symmetry, i.e. $N_D = (N_1, N_2)^T \simeq \mathbf{2} \in G_D$. Since $N_2$ belongs to the dark sector, it carries a nontrivial $G_R$ charge. As a result, neutrino masses receive a type-I seesaw contribution via $N_1$ and a dark radiative contribution mediated by $\eta$ and $N_2$, as depicted in Figure~\ref{o5seesaws}.

The minimal matter content and charge assignments following the assumptions made before are summarized in Table~\ref{tab:1}. In the following sections we explore the two smallest groups that contain a two-dimensional irreducible representation $\bf 2$
$G_D \in \{S_3, D_4\}$ and show that the minimal particle content proposed needs an extra RH neutrino $N_S$, singlet of $G_D$, to reproduce the full oscillation phenomenology, while still requiring fewer fields than a successful $A_4$ realization~\cite{Bonilla:2023pna}.

\begin{table}[h!]
    \centering 
    \begin{tabular}{|c|c|c|c|c|c|c||c|c|}
    \hline
    \text{Group} & $Q_{L_i}$ & $L_i$ & $u_{R_i}$ & $d_{R_i}$ & $\ell_{R_i}$ & $N_D$ & $H_1$ & $\Phi$ \\ \hline\hline
    $SU(2)_L$ & {\bf 2} & {\bf 2} & {\bf 1} & {\bf 1} & {\bf 1} & {\bf 1} & {\bf 2} & {\bf 2}\\
    $G_D$     & {\bf 1} & {\bf 1} & {\bf 1} & {\bf 1} & {\bf 1} & {\bf 2} & {\bf 1} & {\bf 2}\\
    \hline
    \end{tabular}
    \caption{Minimal particle content and charge assignments under a hypothetical $G_D$ non-Abelian discrete symmetry. The additions to the SM are those fields with non-trivial transformation under it, $\Phi = (H_2, \eta)^T$ and $N_D = (N_1, N_2)^T$. The index $i=1,2,3$ labels the three fermion families.}
    \label{tab:1}
\end{table}
\subsection{Constraints}
\label{sec:constraints}

To assess the phenomenological viability of the models constructed in this work, we 
impose constraints from both the lepton and scalar sectors. In the lepton sector, we 
require consistency with the latest neutrino oscillation data from the global fit 
of~\cite{Esteban:2024eli} (see also~\cite{deSalas:2020pgw}), whose $1\sigma$ and 
$3\sigma$ bounds under the assumption of normal ordering (NO) are summarized in 
Table~\ref{tab:nufit}.

\begin{table}[h]
\centering
\begin{tabular}{lccc}
\hline
Parameter & Best fit & $1\sigma$ & $3\sigma$ \\
\hline
$\sin^2\theta_{12}$ & $0.308$ & $0.297 - 0.320$ & $0.275 - 0.345$ \\
$\sin^2\theta_{23}$ & $0.470$ & $0.457 - 0.487$ & $0.435 - 0.585$ \\
$\sin^2\theta_{13}$ & $0.02215$ & $0.02157 - 0.02271$ & $0.02030 - 0.02388$ \\
$\delta_\text{CP}\ [\text{rad}/\pi]$ & $1.18$ & $0.95 - 1.32$ & $0.69 - 2.02$ \\
$\Delta m^2_{21}\ [10^{-5}\ \text{eV}^2]$ & $7.49$ & $7.30 - 7.68$ & $6.92 - 8.05$ \\
$\Delta m^2_{31}\ [10^{-3}\ \text{eV}^2]$ & $2.513$ & $2.494 - 2.534$ & $2.451 - 2.578$ \\
\hline
\end{tabular}
\caption{Experimental bounds on neutrino oscillation observables assuming normal 
ordering (NO). Data taken from the global fit analysis in~\cite{Esteban:2024eli}.}
\label{tab:nufit}
\end{table}

To explore the viable parameter space we perform a random scan subject to the 
following constraints. In the Yukawa sector, all couplings are varied in the range
\begin{equation}
    10^{-10} < |y_i| < \sqrt{4\pi}\,,
    \label{eq:yukawa_range}
\end{equation}
and the masses of the new heavy Majorana fermions are scanned over $[10^2-10^5]\ \text{GeV}$. An additional perturbativity condition arises from the top-quark 
Yukawa coupling with $H_1$. Using $m_t = 172.57$ GeV~\cite{PhysRevLett.132.261902,ParticleDataGroup:2024cfk},
\begin{align}
\label{eq:ytcond}
    y_t = \frac{\sqrt{2}\,m_t}{v_{H_1}}< \sqrt{4\pi},
\end{align}
where $\langle H_1\rangle=v_{H_{1}}$. The condition in Eq.~(\ref{eq:ytcond}) implies $v_{H_1} > 69$ GeV. This can also be satisfied by imposing a bound on
\begin{align}
    \tan\beta\equiv \frac{v_{H_2}}{v_{H_1}} < 3.422.
\end{align}
where $\langle H_2\rangle=v_{H_{2}}$. In the scalar sector, the quartic couplings must satisfy boundedness from below, as 
detailed in Appendix~\ref{app:sca}, and perturbativity, $|\lambda_i| < 4\pi$. To ensure that the couplings of the 125.38 GeV Higgs-like boson of the theory remain consistent with SM Higgs measurements at the LHC~\cite{ParticleDataGroup:2024cfk,ATLAS:2024lyh,CMS:2026nce}, we impose
a mixing between the $Z_2$-even particles to be 
\begin{equation}
    |\theta_H| \lesssim 0.317\,,
    \label{eq:cos2theta}
\end{equation}
which guarantees approximate alignment with the SM expectation.
\\

 At the collider level, the new scalars must satisfy a series of direct-search bounds. For the charged scalars, LEP searches~\cite{ALEPH:2013htx} set a robust lower limit $m_{\eta^\pm},\, m_{h_2^\pm} \gtrsim 100~\text{GeV}$. Stronger but more model-dependent bounds follow from reinterpreting LHC slepton searches. One of these is the ATLAS analysis of~\cite{ATLAS:2014zve}, recast for inert-doublet-like states, which translates into $m_{\eta^\pm} \gtrsim 270~\text{GeV}$ for the dark charged scalars, while a more conservative interpretation~\cite{Karan:2023adm} suggests $m_{h_2^\pm} \gtrsim 400~\text{GeV}$ for the active charged scalars. We adopt these as guideline bounds in the scan. The benchmark points presented in the following sections comfortably satisfy all of them. For the new neutral scalars, ALEPH~\cite{ALEPH:2001roc} together with the di-boson ($WW$, $ZZ$, $\gamma\gamma$) Higgs decay measurements at ATLAS~\cite{ATLAS:2012yxc} and CMS~\cite{CMS:2013zma} imply a lower bound of about $150~\text{GeV}$ on the masses of additional active neutral scalars. For the dark sector, the LEP-based reanalysis of~\cite{Lundstrom:2008ai} excludes the region in which the three conditions $m_{\eta^0} < 80~\text{GeV}$, $m_{\eta^A} < 100~\text{GeV}$, and $m_{\eta^A} - m_{\eta^0} > 8~\text{GeV}$ are simultaneously fulfilled. 
 \\
 
 In addition, when the dark matter candidate is the neutral component of $\eta$, 
the inert-doublet relic-abundance analysis~\cite{Garcia-Cely_2016} singles out 
two viable mass windows,
\begin{equation}
    50~\text{GeV} \lesssim m_{\rm DM} \lesssim 70~\text{GeV} 
    \quad\text{and}\quad 
    535~\text{GeV} \lesssim m_{\rm DM} \lesssim 20~\text{TeV},
    \label{eq:idm-windows}
\end{equation}
in which the correct relic density is reproduced consistently with all collider and 
direct-detection constraints. For masses lying between these two windows, the 
predicted relic abundance \cite{AtacamaCosmologyTelescope:2025blo,Planck:2018vyg} generically falls below the measured value. The benchmark 
points discussed below are chosen so that, whenever the dark matter candidate is 
scalar, its mass falls in the upper window of Eq.~\eqref{eq:idm-windows}. We have 
checked that, in the present framework, the additional portal couplings from the 
extended scalar sector open further annihilation channels beyond those of the pure 
inert doublet model, generically driving the predicted relic abundance below the 
observed value for both the $S_3$ and $D_4$ benchmark points. We refer to 
Ref.~\cite{Garcia-Cely_2016} for a dedicated study of this mass window.
\\

Finally, all Yukawa couplings are taken to be real throughout the scan, so that CP 
is conserved in both the Yukawa and scalar sectors. This is a minimal choice that 
proves sufficient to reproduce all observed oscillation observables under normal 
ordering. Allowing for complex couplings would broaden the parameter space and could 
potentially yield viable solutions for inverted ordering as well, but this 
possibility is left for future work. For each parameter point we compute the 
neutrino mass matrix $M_\nu$, diagonalize both mass matrices, and extract the 
oscillation observables $\sin^2\theta_{ij}$, $\Delta m^2_{21}$, and $\Delta m^2_{31}$. 
A point is deemed viable if all five observables fall within the $3\sigma$ ranges of 
Table~\ref{tab:nufit}.

\section{3HDM with $S_3$ Symmetry}
\label{sec:s3}
According to Table~\ref{tab:1}, the minimal non-Abelian group must have order at
least 6 and admit a two-dimensional irreducible representation. This identifies
$S_3$, the permutation group of three objects.\footnote{$S_3$ has order~6 and
three irreducible representations: $\mathbf{1}$, $\mathbf{1'}$, and $\mathbf{2}$.
Details are given in Appendix~\ref{app:s3d4}.}
\\

The quark sector transforms trivially under $S_3$. The $SU(2)_L$ doublets $L_i$
and the right-handed charged leptons $\ell_{R_j}$ are assigned to the same
singlet representation of $S_3$. With this choice, $M_\ell$ is a non-diagonal 
matrix which gives enough freedom to fit charged lepton masses and whose mixing contributes to or accounts for the lepton mixing matrix $U_{L}=U_\ell^\dagger U_\nu$, where $U_\ell$ diagonalizes the square mass matrix $U_\ell^\dagger M_\ell M_\ell^{\dagger} U_\ell=\text{diag}(m_e^2,m_\mu^2,m_\tau^2)$. We choose all leptons to transform as the trivial
$S_3$ singlet $\mathbf{1}$.
\\

The Majorana mass term of the RH-neutrino doublet is
\begin{align}
    -\mathcal{L} \supset \frac{m_N}{2}\overline{N^c_D}N_D + h.c.
    = \frac{m_N}{2}(\overline{N^c_1}N_1 + \overline{N^c_2}N_2) + h.c.,
    \label{NDM}
\end{align}
giving two degenerate, unmixed RH neutrinos.  The charged-lepton Yukawa interactions are given by, 
\begin{align}
    -\mathcal{L} \supset \sum_{i,j} y_{ij}\overline{L_i}H_1\ell_{R_j} + h.c.
    \qquad
\end{align}
After electroweak symmetry breaking the mass matrix for charged leptons turns out to be
\begin{equation}
 \label{eq:MlS3}
     M_\ell = \frac{v_{H_1}}{\sqrt{2}}\begin{pmatrix}
        y_{ee} & y_{e\mu} & y_{e\tau} \\
        y_{\mu e} & y_{\mu\mu} & y_{\mu\tau} \\
        y_{\tau e} & y_{\tau\mu} & y_{\tau\tau}
    \end{pmatrix},
\end{equation}
where $\langle H_1 \rangle=v_{H_1}$ is the vev of $H_1$. The scalar potential is detailed in
Appendix~\ref{app:sca}.
The invariant Lagrangian for the neutrino sector is 
\begin{align}
    -\mathcal{L}_{\nu} = \sum_{i} y^D_{i}\overline{L}_i
    (\tilde{\Phi} N_D)_{\mathbf{1}} + h.c.
    = \sum_{i} y^D_{i}
    (\overline{L}_i \tilde{H}_2 N_1 + \overline{L}_i \tilde{\eta} N_2) + h.c.
\end{align}
where $\tilde{\varphi}=i\sigma_2 \varphi^*$ (with $\varphi=H_2,\eta$). 
These interactions, together with Eq.~(\ref{NDM}), generate the neutrino mass matrix at tree level and one loop,

\begin{align}
    M_{\nu}^D = \left(\frac{v_{H_2}^2}{2m_N} + C_1 + C_2\right)
    \begin{pmatrix}
        (y^D_e)^2 & y^D_e y^D_\mu & y^D_e y^D_\tau \\
        y^D_e y^D_\mu & (y^D_\mu)^2 & y^D_\mu y^D_\tau \\
        y^D_e y^D_\tau & y^D_\mu y^D_\tau & (y^D_\tau)^2
    \end{pmatrix},
\end{align}
\\

where $\langle\Phi\rangle=(\langle H_2\rangle,\langle\eta\rangle)^T=(v_{H_2},0)^T$, $C_1$ and $C_2$ are the loop functions defined in Appendix~\ref{app:loop}. This matrix is \emph{rank-1} regardless of the values of the couplings. A rank-1 matrix produces only one non-zero mass
eigenvalue and cannot simultaneously account for $\Delta m^2_{\rm atm}$ and
$\Delta m^2_{\rm sol}$.
\\

The natural remedy is to introduce an additional singlet RH neutrino transforming trivially under $S_3$, i.e. $N_S\sim\bf{1}$. The charge assignments of the model are summarized in Table~\ref{tab:nmmcs3}. 
\\

\begin{table}[h!]
    \centering 
    \begin{tabular}{|c|c|c|c||c|c|c|}
    \hline
    \text{Group} & $L_i$ & $\ell_{R_i}$ & $H_1$ & $N_S$ & $N_D$ & $\Phi$ \\ \hline\hline
    $SU(2)_L$ & {\bf 2} & {\bf 1} & {\bf 2} & {\bf 1} & {\bf 1} & {\bf 2}\\
    $S_3$ & {\bf 1\,} & {\bf 1\,} & {\bf 1} & {\bf 1\,} & {\bf 2} & {\bf 2}\\
    \hline
    \end{tabular}
    \caption{Particle content and charge assignments for the viable $S_3$
    model, with $i=e,\mu,\tau$. Both $N_S$ and the lepton fields must transform 
    as the same singlet type.}
    \label{tab:nmmcs3}
\end{table}

Then the neutrino Lagrangian has these new contributions,
\begin{align}
    -\mathcal{L}_\nu \supset \sum_{i} y^S_i \overline{L_i} H_1 N_S + \frac{m_S}{2}\overline{N^c_S}N_S + h.c.
    \label{NSM}
\end{align}
Then the tree-level and one-loop contributions to the neutrino mass matrix from $N_S$ is given by,
\begin{align}
    M_\nu^S=\left(\frac{v_{H_1}^2}{2m_S} + C_3\right)
    \begin{pmatrix}
        (y^S_e)^2 & y^S_e y^S_\mu & y^S_e y^S_\tau \\
        y^S_e y^S_\mu & (y^S_\mu)^2 & y^S_\mu y^S_\tau \\
        y^S_e y^S_\tau & y^S_\mu y^S_\tau & (y^S_\tau)^2
    \end{pmatrix},
\end{align}
where $C_3$ is defined in Appendix~\ref{app:loop}. Since the Yukawa vectors
$\mathbf{y}^D$ and $\mathbf{y}^S$ are independent, the total mass matrix is
\begin{align}
\label{eq:MnuS3r2}
     \\
    M_{\nu}&= \left(\frac{v_{H_2}^2}{2m_N} + C_1 + C_2\right)
    \begin{pmatrix}
        (y^D_e)^2 & y^D_e y^D_\mu & y^D_e y^D_\tau \\
        y^D_e y^D_\mu & (y^D_\mu)^2 & y^D_\mu y^D_\tau \\
        y^D_e y^D_\tau & y^D_\mu y^D_\tau & (y^D_\tau)^2
    \end{pmatrix} \nonumber + \left(\frac{v_{H_1}^2}{2m_S} + C_3\right)
    \begin{pmatrix}
        (y^S_e)^2 & y^S_e y^S_\mu & y^S_e y^S_\tau \\
        y^S_e y^S_\mu & (y^S_\mu)^2 & y^S_\mu y^S_\tau \\
        y^S_e y^S_\tau & y^S_\mu y^S_\tau & (y^S_\tau)^2
    \end{pmatrix}.
\end{align}

The resulting total neutrino mass matrix in Eq.~(\ref{eq:MnuS3r2}) is generically \emph{rank-2}, yielding one massless neutrino and two massive eigenstates, sufficient to accommodate both mass-squared differences simultaneously. The structures of $M_\ell$, Eq.~(\ref{eq:MlS3}), and  $M_\nu$, Eq.~(\ref{eq:MnuS3r2}), provide sufficient parametric freedom to account for the leptonic mixing angles $U_L$. In particular, the rank-2 structure of $M_\nu$ enforces a vanishing lightest neutrino mass, $m_{\nu_1} = 0$, so that the spectrum is fully determined by the two measured mass-squared differences as $m_{\nu_2} = \sqrt{\Delta m^2_{21}}$ and $m_{\nu_3} = \sqrt{\Delta m^2_{31}}$ in normal ordering. Other possible charge assignments, as $S_3$ singlets, for charged leptons generate non-diagonal rank-3 mass matrices with some null elements. Even though it is possible to generate a \emph{rank-2} neutrino mass matrix, none of these combinations can fit the observed mixing angles. The assignment with $N_S \sim \mathbf{1'}$
instead forces the two rank-1 contributions to be collinear, keeping the matrix at rank-1.
Extensions that add further scalar doublets are not considered, as they exceed the 3HDM
framework.
\\

As a direct phenomenological consequence of $m_{\nu_1}=0$, the effective Majorana mass 
governing neutrinoless double beta decay~\cite{Rodejohann:2011vc},
\begin{equation}
    \langle m_{\beta\beta} \rangle = \left| \sum_j U_{ej}^2\, m_{\nu_j} \right|
    = \left| c_{12}^2 c_{13}^2\, m_{\nu_1} + s_{12}^2 c_{13}^2\, m_{\nu_2}\, e^{2i\phi_{12}} + s_{13}^2\, m_{\nu_3}\, e^{2i\phi_{13}} \right|,
    \label{eq:mbb}
\end{equation}
reduces in the present model to a sum of only the $m_{\nu_2}$ and $m_{\nu_3}$ 
contributions. Since the model conserves CP in both the Yukawa and scalar sectors, 
the Majorana phases $\phi_{12}, \phi_{13}$ are discrete, taking values $0$ or 
$\pi/2$, so that the relative signs of the two surviving terms are $\pm 1$. Using 
the benchmark values of Table~\ref{tab:S3_output}, this yields the prediction
\begin{equation}
    \langle m_{\beta\beta} \rangle^{S_3} \simeq 1.45\text{--}3.70~\text{meV},
\end{equation}
where the lower (upper) value corresponds to a destructive (constructive) 
interference between the two contributions, well below the reach of current 
neutrinoless double beta decay searches. The strongest such bound to date comes 
from the KamLAND-Zen 800 experiment~\cite{kamLand}, based on the full $^{136}$Xe 
dataset, which improves on the results from AMoRE-I~\cite{AMORE} and 
CUPID-Mo~\cite{cupid-mo}, and sets
\begin{equation}
\label{0nubb-limit}
    \langle m_{\beta\beta} \rangle < 28\text{--}122~\text{meV},
\end{equation}
more than an order of magnitude above the present prediction. Next-generation experiments will be needed to probe this region of parameter space.
\\

To demonstrate the numerical viability of this $S_3$ model in its unique viable assignment up to equivalence (leptons as $\mathbf{1}$, $N_S \sim \mathbf{1}$). The alternative choice (leptons as $\mathbf{1'}$, $N_S \sim \mathbf{1'}$) leads to identical phenomenology, since $\mathbf{1'}\otimes\mathbf{1'} = \mathbf{1}$ yields the same invariant Yukawa structures, we present the benchmark point given in Table~\ref{tab:S3_input}, consistent with normal ordering (NO) and the global-fit values of Table~\ref{tab:nufit}. The resulting neutrino oscillation observables, collected in Table~\ref{tab:S3_output}, 
show excellent agreement with the global fits. The dark matter candidate in this scenario 
is the CP-even scalar $\eta^0$, with $m_{\eta^0}=540.9$~GeV, safely above the 400~GeV 
threshold set by the collider constraints of Section~\ref{sec:constraints}.
We have checked that the relic abundance and 
direct-detection prospects of this dark matter candidate are consistent with 
the dedicated analysis of this mass window in Ref.~\cite{Garcia-Cely_2016}.

\begin{table}[h!]
\centering
\begin{tabular}{llll}
\hline
\multicolumn{4}{c}{\textbf{Input Parameters}} \\
\hline
$y^D_e$    & $-2.33273\times10^{-6}$ & $y_{ee}$   & $\phantom{-}1.876449333333\times10^{-2}$ \\
$y^D_\mu$  & $\phantom{-}3.0548\times10^{-6}$ & $y_{\mu\mu}$ & $\phantom{-}1.116477917\times10^{-3}$ \\
$y^D_\tau$ & $\phantom{-}2.19097\times10^{-7}$ & $y_{\tau\tau}$ & $\phantom{-}5.4839089518333\times10^{-6}$ \\
$y^S_e$    & $\phantom{-}3.5532\times10^{-6}$ & $y_{e\mu}$   & $\phantom{-}1.4014590666\times10^{-6}$ \\
$y^S_\mu$  & $-7.18242\times10^{-9}$ & $y_{e\tau}$  & $\phantom{-}1.650424491333\times10^{-4}$ \\
$y^S_\tau$ & $-2.06695\times10^{-6}$ & $y_{\mu\tau}$ & $-1.647218325\times10^{-5}$ \\
$m_N$ [GeV] & $\phantom{-}7.051\times10^{3}$     & $y_{\mu e}$  & $-5.03160733333\times10^{-4}$ \\
$m_S$ [GeV] & $\phantom{-}1.085\times10^{4}$     & $y_{\tau e}$ & $\phantom{-}2.01404216473\times10^{-7}$ \\
&  & $y_{\tau\mu}$ & $-7.70627513657\times10^{-6}$  \\
\hline
$\lambda_1$ & $\phantom{-}0.0011596$ &
$\lambda_2$ & $\phantom{-}10.40664$ \\
$\lambda_3$ & $-3.77414$ &
$\lambda_4$ & $\phantom{-}3.70955$ \\
$\lambda_5$ & $-4.49015$ &
$\lambda_6$ & $-1.21839$ \\
$\lambda^{S_3}_7$ & $-1.02907$ &
$\lambda^{S_3}_8$ & $\phantom{-}4.7056$ \\
$t_\beta$   & $\phantom{-}1.54172474037$ & $|\theta_h|$ [rad] & $\phantom{-}0.291$ \\
\hline
\end{tabular}
\caption{Input parameters for the $S_3$ next-to-minimal benchmark (NO, $\mathbf{1,1,1}$ lepton assignment).}
\label{tab:S3_input}
\end{table}

\begin{table}[h!]
\centering
\begin{tabular}{lllll}
\hline
\multicolumn{5}{c}{\textbf{Output Parameters}} \\
\hline
\multicolumn{5}{c}{\textit{Scalar mass spectrum}} \\
\hline
$m_{h^0_1}$ [GeV] & $125.2$ &\quad& $m_{\eta^0}$ [GeV] & $540.9$ \\
$m_{h^0_2}$ [GeV] & $873.5$ &\quad& $m_{\eta^A}$ [GeV] & $770.4$ \\
$m_{h^A_2}$ [GeV] & $428.4$ &\quad& $m_{\eta^\pm}$ [GeV] & $548.9$ \\
$m_{h^\pm_2}$ [GeV] & $531.5$ &\quad&  &  \\
\hline
\multicolumn{5}{c}{\textit{Neutrino oscillation observables}} \\
\hline
$\sin^2\theta_{12}$ & $0.3043$ &\quad& $\Delta m^2_{21}\ [10^{-5}\ \text{eV}^2]$ & $7.483$ \\
$\sin^2\theta_{13}$ & $0.02250$ &\quad& $\Delta m^2_{31}\ [10^{-3}\ \text{eV}^2]$ & $2.507$ \\
$\sin^2\theta_{23}$ & $0.4742$ & & & \\
\hline
\end{tabular}
\caption{Output parameters for the $S_3$ benchmark, scalar mass spectrum and neutrino
oscillation observables. All mixing angles and mass splittings lie within $1\sigma$ of
global fits~\cite{deSalas:2020pgw,Esteban:2024eli}.}
\label{tab:S3_output}
\end{table}


\section{3HDM with $D_4$ Symmetry}
\label{sec:d4}

The second smallest non-Abelian group with a doublet representation is $D_4$,
the symmetry group of a square with order 8 and five irreducible representations, four one-dimensional $\bf{1}_{++}$, $\bf{1}_{--}$, $\bf{1}_{+-}$, $\bf{1}_{-+}$, and one two-dimensional $\bf{2}$. Details are given in Appendix~\ref{app:s3d4}.
\\

Following the recipe described in the previous section, we construct viable models 
using the $D_4$ symmetry group. As before, the quark sector is assumed to be blind 
under the additional global symmetry. For leptons, it is possible to recover Yukawa structures similar to those obtained with the $S_3$ group. In the minimal case, with an extra $D_4$ scalar doublet $\Phi\sim\bf{2}$ and a Majorana fermion $N_D\sim\bf{2}$, this happens when the left and right-handed charged leptons transform as the same $D_4$ singlet, and the same holds when an extra fermion $N_S$ with a similar transformation is added. Other charge assignments for the $D_4$ leptons, with or without $N_S$, lead to a diagonal squared mass matrix for the charged leptons, i.e. $M_\ell M_\ell^\dagger=\text{diag}(m_e^2,m_\mu^2,m_\tau^2)$ with $U_\ell=\mathbb{I}$. However, the resulting neutrino mass matrix in these cases does not account for the oscillation data.
\\

The only $D_4$-symmetric scenarios inequivalent to those arising in $S_3$ models are those 
in which the charged-lepton squared mass matrix $M_\ell M_\ell^\dagger$ exhibits one of the 
following textures,
\begin{equation}
     M_\ell M_\ell^\dagger \sim
     \begin{pmatrix}
        \times & 0 & 0 \\
        0 & \times & \times \\
        0 & \times & \times
    \end{pmatrix},
     \ \ 
     \begin{pmatrix}
        \times & 0 & \times \\
        0 & \times & 0 \\
        \times & 0 & \times
    \end{pmatrix},
    \ \
     \begin{pmatrix}
        \times & \times & 0 \\
        \times & \times & 0 \\
        0 & 0 & \times
    \end{pmatrix},
    \label{eq:ML2structures}
\end{equation}
where $\times$ denotes a generically non-zero entry.  Among the charge assignments that lead to these textures, some fail to generate mixing among all three lepton flavors and are therefore phenomenologically excluded, while others do produce the required three-flavor mixing. In what follows, we present in detail one representative example of the latter, corresponding to the first texture in Eq.~\eqref{eq:ML2structures}, illustrating how the structure arises from a concrete $D_4$ assignment. The remaining two textures admit analogous viable realizations. They are obtained by singling out the muon or the tau lepton doublet with a distinct $D_4$ representation, which we have verified yield equally good fits to oscillation data, with 
phenomenology indistinguishable from the representative case discussed below. In contrast to the $S_3$ scenario discussed in Section~\ref{sec:s3}, where the rank-2 structure of $M_\nu$ enforced a vanishing lightest neutrino mass, all three viable $D_4$ assignments yield a \emph{rank-3} neutrino mass matrix. Consequently, the three light neutrinos are \emph{all massive}, with $m_{\nu_1}\neq 0$ being a model-dependent prediction rather than a fixed value. The effective Majorana mass $\langle m_{\beta\beta}\rangle$, defined in Eq.~\eqref{eq:mbb}, therefore receives a non-vanishing contribution from $m_{\nu_1}$ as well.

\subsubsection*{Electron transforming as $\mathbf{1_{+-}}$ (representative example)}

The first structure in Eq.~\eqref{eq:ML2structures} is realized when the electron 
lepton doublet $L_e$ transforms differently from $L_{\mu,\tau}$ under $D_4$, as shown 
in Table~\ref{tab:nmmcd4E}. The remaining two inequivalent cases, corresponding to the second and third 
textures in Eq.~\eqref{eq:ML2structures}, arise when either $L_\mu$ or $L_\tau$ is 
singled out with a distinct $D_4$ representation. Both admit equally viable fits 
and are not discussed further.

\begin{table}[h!]
    \centering 
    \begin{tabular}{|c|c|c|c|c|c||c|c|c|}
    \hline
    \text{Group} & $L_e$ & $L_{\mu,\tau}$ & $\ell_{R_e}$ & $\ell_{R_{\mu,\tau}}$ & $H_1$ & $N_S$ & $N_D$ & $\Phi$ \\ \hline\hline
    $SU(2)_L$ & {\bf 2} & {\bf 2} & {\bf 1} & {\bf 1} & {\bf 2} & {\bf 1} & {\bf 1} & {\bf 2}\\
    $D_4$ & {\bf 1$_{+-}$} & {\bf 1$_{++}$} & {\bf 1$_{+-}$} & {\bf 1$_{++}$} & {\bf 1$_{++}$} & {\bf 1$_{++}$} & {\bf 2} & {\bf 2}\\
    \hline
    \end{tabular}
    \caption{Particle content and charge assignments for the next-to-minimal $D_4$
    model. $N_S$ carries the same discrete charge as $L_{\mu,\tau}$.}
    \label{tab:nmmcd4E}
\end{table}

The most general Yukawa Lagrangian invariant under $SU(2)_L \times D_4$ consistent 
with the charge assignments of Table~\ref{tab:nmmcd4E} reads
\begin{align}
    -\mathcal{L} \supset y_{ee}\overline{L_e}H_1\ell_{R_e}
    + y^D_e\overline{L_e}(\tilde{\Phi}N_D)_{+-} + \sum_{i,j} y_{ij}\overline{L_i}H_1\ell_{R_j} 
    + y^D_i\overline{L_i}(\tilde{\Phi}N_D)_{++} + y^S_{i}\overline{L_i}\tilde{H}_1N_S + \text{h.c.},
\end{align}
where $i,j = \mu, \tau$. One can verify that each term is a singlet under $D_4$, 
ensuring full invariance of the Lagrangian under all imposed symmetries. Upon electroweak 
symmetry breaking, the charged-lepton and neutrino mass matrices take the form
\begin{align}
        M_\ell =&~ \dfrac{v_{H_1}}{\sqrt{2}}\begin{pmatrix}
         y_{ee} & 0 & 0 \\
         0 &  y_{\mu\mu} & y_{\mu\tau} \\
         0 & y_{\tau\mu} &  y_{\tau\tau} 
        \end{pmatrix},
\end{align}
\begin{align}
     M_{\nu} =&~ \left(\dfrac{v_{H_{2}}^2}{2m_N}+C_1\right)\begin{pmatrix}
        (y^D_{e})^2 & y^D_{e}y^D_{\mu} & y^D_{e}y^D_{\tau} \\
        y^D_{e}y^D_{\mu} &  (y^D_{\mu})^2 &  y^D_{\mu}y^D_{\tau} \\
         y^D_{e}y^D_{\tau} &   y^D_{\mu}y^D_{\tau} &  (y^D_{\tau})^2 
        \end{pmatrix} + C_2\begin{pmatrix}
        (y^D_{e})^2 & -y^D_{e}y^D_{\mu} & -y^D_{e}y^D_{\tau} \\
        -y^D_{e}y^D_{\mu} &  (y^D_{\mu})^2 &  y^D_{\mu}y^D_{\tau} \\
         -y^D_{e}y^D_{\tau} &   y^D_{\mu}y^D_{\tau} &  (y^D_{\tau})^2 
        \end{pmatrix} \nonumber \\
        &~+\left(\dfrac{v_{H_{1}}^2}{2m_S}+C_3\right)\begin{pmatrix}
        0 & 0 & 0 \\
        0 &  (y^S_{\mu})^2 &  y^S_{\mu}y^S_{\tau} \\
         0 &   y^S_{\mu}y^S_{\tau} &  (y^S_{\tau})^2 
        \end{pmatrix},
\end{align}
where $M_\ell M_\ell^\dagger$ reproduces the first texture in 
Eq.~\eqref{eq:ML2structures}, as expected from the symmetry assignments. A representative benchmark point consistent with normal ordering (NO) is summarized in Tables~\ref{tab:D4_input_e} and~\ref{tab:D4_output_e}, illustrating the numerical viability of this scenario. The chosen point fulfills all theoretical and collider bounds of Section~\ref{sec:constraints}, with every new scalar mass exceeding 400~GeV. Here, the dark matter candidate corresponds to the CP-even scalar $\eta^0$, with a mass of $549.4$~GeV. The neutrino oscillation observables reported in Table~\ref{tab:D4_output_e} show excellent agreement with the global fits of 
Table~\ref{tab:nufit}.
\\

\begin{table}[h!]
\centering
\begin{tabular}{llll}
\hline
\multicolumn{4}{c}{\textbf{Input Parameters} ($e$ transforms differently)} \\
\hline
$y^D_e$  & $-2.61387\times10^{-6}$ & $y_{ee}$     & $\phantom{-}5.49373069\times10^{-6}$ \\
$y^D_\mu$  & $\phantom{-}5.35132\times10^{-6}$ & $y_{\mu\mu}$  & $\phantom{-}1.13592768\times10^{-3}$ \\
$y^D_\tau$ & $\phantom{-}1.65617\times10^{-6}$ & $y_{\tau\tau}$ & $-1.9102929\times10^{-2}$ \\
$y^S_\mu$  & $-2.93929\times10^{-6}$ & $y_{\mu\tau}$ & $\phantom{-}5.3074811455\times10^{-6}$ \\
$y^S_\tau$ & $-4.93562\times10^{-6}$ & $y_{\tau\mu}$ & $\phantom{-}1.4976997\times10^{-4}$ \\
$m_N$ [GeV]& $\phantom{-}3.9393\times10^{4}$ & & \\
$m_S$ [GeV]& $\phantom{-}7.558\times10^{3}$ & & \\
\hline
$\lambda_1$ & $\phantom{-}0.867572$ &
$\lambda_2$ & $\phantom{-}6.09711$ \\
$\lambda_3$ & $\phantom{-}0.271674$ &
$\lambda_4$ & $\phantom{-}5.97418$ \\
$\lambda_5$ & $-5.0378$ &
$\lambda_6$ & $-3.01432$ \\
$\lambda^{D_4}_7$ & $-3.31231$ &
$\lambda^{D_4}_8$ & $\phantom{-}0.180202$ \\
$t_\beta$   & $\phantom{-}1.58029$ & $|\theta_h|$ [rad] & $\phantom{-}0.247$ \\
\hline
\end{tabular}
\caption{Input parameters for the $D_4$ benchmark with $L_e$ transforming differently 
from $L_\mu$ and $L_\tau$.}
\label{tab:D4_input_e}
\end{table}

\begin{table}[h!]
\centering
\begin{tabular}{lllll}
\hline
\multicolumn{5}{c}{\textbf{Output Parameters} ($e$ transforms differently)} \\
\hline
\multicolumn{5}{c}{\textit{Scalar mass spectrum}} \\
\hline
$m_{h^0_1}$ [GeV] & $125.2$ &\quad& $m_{\eta^0}$ [GeV] & $549.4$ \\
$m_{h^0_2}$ [GeV] & $505$ &\quad& $m_{\eta^A}$ [GeV] & $561.2$ \\
$m_{h^A_2}$ [GeV] & $427.1$ &\quad& $m_{\eta^\pm}$ [GeV] & $596.6$ \\
$m_{h^\pm_2}$ [GeV] & $493.6$ &\quad&  &  \\
\hline
\multicolumn{5}{c}{\textit{Neutrino oscillation observables}} \\
\hline
$\sin^2\theta_{12}$ & $0.3116$ &\quad& $\Delta m^2_{21}\ [10^{-5}\ \text{eV}^2]$ & $7.472$ \\
$\sin^2\theta_{13}$ & $0.02223$ &\quad& $\Delta m^2_{31}\ [10^{-3}\ \text{eV}^2]$ & $2.509$ \\
$\sin^2\theta_{23}$ & $0.4695$ &  &  &  \\
\hline
\end{tabular}
\caption{Output parameters for the $D_4$ benchmark ($L_e$ different), showing the scalar mass spectrum
and neutrino oscillation observables. All mixing angles and mass splittings lie within $1\sigma$ of
current global fits shown in Table~\ref{tab:nufit}.}
\label{tab:D4_output_e}
\end{table}

Since the neutrino mass matrix is rank-3, all three light eigenvalues are non-vanishing and the effective Majorana mass~\eqref{eq:mbb} receives contributions from the full spectrum. Using the oscillation observables of Table~\ref{tab:D4_output_e} together with the CP-conserving choices for the Majorana phases ($\phi_{12}, \phi_{13} \in \{0,\pi/2\}$), the prediction for this benchmark spans the indicative range
\begin{equation}
    \langle m_{\beta\beta} \rangle^{D_4} \simeq 1\text{--}8~\text{meV}\quad \text{for}\quad m_{\nu_1} \lesssim 5~\text{meV},
\end{equation}
 with the precise value within this band fixed by $m_{\nu_1}$ and the relative CP signs determined dynamically by the Yukawa structure. We have checked that the two remaining textures of Eq.~\eqref{eq:ML2structures}, those obtained by singling out $L_\mu$ or $L_\tau$ instead of $L_e$, also admit viable benchmark points, with oscillation observables and $\langle m_{\beta\beta}\rangle$ predictions essentially 
indistinguishable from the representative case shown here. We therefore do not display them explicitly. All values are well below the present sensitivity of the most stringent $0\nu\beta\beta$ searches, Eq.~\eqref{0nubb-limit}.


\section{Conclusions}
\label{sec:conc}

We have constructed and analyzed minimal three-Higgs-doublet models (3HDMs) realizing the discrete dark matter mechanism within the 
scoto-seesaw framework, using the smallest non-Abelian discrete groups admitting a two-dimensional irreducible representation, $S_3$ (order~6) and $D_4$ (order~8).
\\

It was found that the unique minimal extension of the SM imposing either $S_3$ or $D_4$ as global symmetry, adds two scalar doublets $\Phi$ and three heavy Majorana fermions $N_D$ and $N_S$. In the $S_3$ case, $N_S$ adds an independent rank-1 contribution that lifts the total mass matrix to rank-2. As a direct consequence, the lightest neutrino is exactly massless, 
$m_{\nu_1} = 0$, and the spectrum is fully fixed by the two measured 
mass-squared differences. Hence the model predicts 
$\langle m_{\beta\beta}\rangle^{S_3} \simeq 1.45$--$3.70~\text{meV}$, with the band generated by the discrete CP-conserving choices of the Majorana phases. In the $D_4$ case, $N_S$ must couple to two lepton doublets, generating a third contribution misaligned with the $N_D$ terms and raising the matrix to rank-3 with no structural zeros in $U_L$. Then the three light neutrinos are then all massive, with $m_{\nu_1}\neq 0$ a model-dependent prediction. 
The corresponding $\langle m_{\beta\beta}\rangle$ predictions across the three viable $D_4$ benchmarks are essentially identical, of order 
$1$--$8~\text{meV}$ for $m_{\nu_1} \lesssim 5~\text{meV}$, well below current $0\nu\beta\beta$ bounds but within the reach of next-generation experiments. Through a numerical scan we have verified that both next-to-minimal models reproduce all five oscillation observables within $1\sigma$ of the central values from global fits~\cite{deSalas:2020pgw,Esteban:2024eli}.

The 3HDMs achieve what the $A_4$ realization of~\cite{Bonilla:2023pna} requires a four-Higgs-doublet model to accomplish, reducing the scalar sector by one doublet. A notable distinction is that the 
present models conserve CP in both the Yukawa and scalar sectors, whereas the $A_4$ realization requires CP violation in the scalar potential to reproduce the observed mixing angles.

The dark matter candidate is the lightest $Z_2$-odd particle. In the $S_3$ and $D_4$ benchmarks presented in Sections~\ref{sec:s3} and~\ref{sec:d4}, this role is played by the neutral component of the inert doublet ($\eta^0$ or $\eta^A$), with 
masses in the upper inert-doublet window of Eq.~\eqref{eq:idm-windows}, and phenomenology analogous to that of the well-studied Inert Doublet Model~\cite{Deshpande:1977rw,Barbieri:2006dq,LopezHonorez:2006gr,LopezHonorez:2010tb,Goudelis:2013uca}. 
More generally, the framework also accommodates the heavy Majorana fermion 
$N_2$ from the $D_4$-doublet as the lightest $Z_2$-odd state. A fermionic WIMP with mass near the electroweak scale, however, is generically expected to be \emph{overabundant}, since the same Yukawa couplings that must remain small to reproduce the observed neutrino 
masses also suppress the annihilation cross 
section~\cite{Merle:2016scw,Karan:2023adm}. A dedicated study of the dark 
matter parameter space, including relic density and direct 
detection constraints, is deferred to future work.

\section*{Acknowledgments}
Work supported by ANID-Chile under the grant ANID FONDECYT/Regular 1241855. C.B acknowledges hospitality and 
support from the ICTP through the Associates Programme (2023-2028). A.L. thanks the support from ANID-Chile through the National Doctoral Fellowship No. 21262169.
The authors acknowledge scientific support provided by NISEU-UCN, Resolución VRIDT N°200/2025.
\appendix

\section{Discrete Non-Abelian Groups}
\label{app:s3d4}

\subsection{$S_3$ Group}

The symmetry group $S_3$ is defined by two generators:
\begin{align}
    S_3 := \langle S,T :~ S^2=T^3=(ST)^2=e \rangle,
\end{align}
where $S$ and $T$ represent a reflection and $2\pi/3$-rotation. It is a
non-Abelian group with six elements and three irreducible representations,
$\mathbf{1}\oplus\mathbf{1'}\oplus\mathbf{2}$. In the $S$-diagonal basis the
generators of the doublet representation take the
form~\cite{Ishimori:2010au}:
\begin{align}
    2 \rightarrow S=\begin{pmatrix}
        1 & 0 \\ 0 & -1
    \end{pmatrix}, \quad T=\begin{pmatrix}
        -1/2 & -\sqrt{3}/2 \\ \sqrt{3}/2 & -1/2
    \end{pmatrix}.
    \label{s3z2}
\end{align}
The corresponding inner products are~\cite{Ishimori:2010au}:
\begin{align}
    x_{\mathbf{1}}\otimes y_{\mathbf{1}} = x_{\mathbf{1'}}\otimes y_{\mathbf{1'}}
    = (xy)_{\mathbf{1}}, \qquad
    x_{\mathbf{1}}\otimes y_{\mathbf{1'}} = (xy)_{\mathbf{1'}},
\end{align}
\begin{align}
    x_{\mathbf{1}}\otimes \begin{pmatrix} y_1 \\ y_2 \end{pmatrix}_{\mathbf{2}}
    = \begin{pmatrix} xy_1 \\ xy_2 \end{pmatrix}_{\mathbf{2}}, \qquad
    x_{\mathbf{1'}}\otimes \begin{pmatrix} y_1 \\ y_2 \end{pmatrix}_{\mathbf{2}}
    = \begin{pmatrix} -xy_2 \\ xy_1 \end{pmatrix}_{\mathbf{2}},
\end{align}
\begin{align}
    \begin{pmatrix} x_1 \\ x_2 \end{pmatrix}_{\mathbf{2}}
    \otimes \begin{pmatrix} y_1 \\ y_2 \end{pmatrix}_{\mathbf{2}}
    = (x_1 y_1 + x_2 y_2)_{\mathbf{1}}
    \oplus (x_1 y_2 - x_2 y_1)_{\mathbf{1'}}
    \oplus \begin{pmatrix} x_2 y_2 - x_1 y_1 \\ x_1 y_2 + x_2 y_1 \end{pmatrix}_{\mathbf{2}}.
\end{align}

\subsection{$D_4$ Group}

The symmetry group $D_4$ is defined by:
\begin{align}
    D_4 := \langle S,T :~ S^2=T^4=e,\; STS=T^{-1} \rangle.
\end{align}
It has eight elements and five irreducible representations,
$\mathbf{1}_{++}\oplus\mathbf{1}_{+-}\oplus\mathbf{1}_{-+}
\oplus\mathbf{1}_{--}\oplus\mathbf{2}$. In the $S$-diagonal
basis~\cite{Ishimori:2010au}:
\begin{align}
    2 \rightarrow S=\begin{pmatrix}
        1 & 0 \\ 0 & -1
    \end{pmatrix}, \quad T=\begin{pmatrix}
        0 & -1 \\ 1 & 0
    \end{pmatrix}.
\end{align}
The inner products are:
\begin{align}
x_{++}\otimes y_{++} = x_{--}\otimes y_{--} =
x_{+-}\otimes y_{+-} = x_{-+}\otimes y_{-+} = (xy)_{++},
\end{align}
\begin{align}
x_{++}\otimes y_{--} = x_{-+}\otimes y_{+-} = (xy)_{--}, \qquad
x_{++}\otimes y_{+-} = x_{--}\otimes y_{-+} = (xy)_{-+},
\end{align}
\begin{align}
x_{++}\otimes y_{-+} = x_{--}\otimes y_{+-} = (xy)_{-+},
\end{align}
\begin{align}
x_{++}\otimes \begin{pmatrix} y_1 \\ y_2 \end{pmatrix}_{\mathbf{2}}
= \begin{pmatrix} xy_1 \\ xy_2 \end{pmatrix}_{\mathbf{2}}, \qquad
x_{+-}\otimes \begin{pmatrix} y_1 \\ y_2 \end{pmatrix}_{\mathbf{2}}
= \begin{pmatrix} xy_1 \\ -xy_2 \end{pmatrix}_{\mathbf{2}},
\end{align}
\begin{align}
x_{-+}\otimes \begin{pmatrix} y_1 \\ y_2 \end{pmatrix}_{\mathbf{2}}
= \begin{pmatrix} xy_2 \\ -xy_1 \end{pmatrix}_{\mathbf{2}}, \qquad
x_{--}\otimes \begin{pmatrix} y_1 \\ y_2 \end{pmatrix}_{\mathbf{2}}
= \begin{pmatrix} xy_2 \\ xy_1 \end{pmatrix}_{\mathbf{2}},
\end{align}
\begin{align}
    \begin{pmatrix} x_1 \\ x_2 \end{pmatrix}_{\mathbf{2}}
    \otimes \begin{pmatrix} y_1 \\ y_2 \end{pmatrix}_{\mathbf{2}}
    = (x_1 y_1 \pm x_2 y_2)_{+\pm} \oplus (x_1 y_2 \mp x_2 y_1)_{-\pm}.
\end{align}

\section{Scalar Sectors}
\label{app:sca}

All models presented here share a common extended scalar sector consisting of a 
$G_D$-doublet $\Phi=(H_2,\eta)^T$ and an $SU(2)_L$ doublet $H_1$ that transforms trivially under 
$G_D \in \{S_3, D_4\}$. The scalar potential is accordingly built upon a common 
set of terms $V_0$, which reads
\begin{align}
V_0 =&~ -\mu^2_H H_1^\dagger H_1 - \mu^2_\Phi(\Phi^\dagger\Phi)_{\mathbf{r}_e}
+ \lambda_1(H_1^\dagger H_1)^2 + \lambda_2(\Phi^\dagger\Phi)_{\mathbf{r}_e}^2 
+ \lambda_3(\Phi^\dagger\Phi)_{\mathbf{r}_a}^2 \nonumber\\
&~ + \lambda_4(H_1^\dagger H_1)(\Phi^\dagger\Phi)_{\mathbf{r}_e}
+ \lambda_5\left[(\Phi^\dagger H_1)_{\mathbf{2}}(H_1^\dagger\Phi)_{\mathbf{2}}\right]_{\mathbf{r}_e}
+ \frac{\lambda_6}{2}\left(\left[(\Phi^\dagger H_1)_{\mathbf{2}}(\Phi^\dagger H_1)_{\mathbf{2}}\right]_{\mathbf{r}_e}
+ \text{h.c.}\right),
\end{align}
where $\mathbf{r}_e = \mathbf{1} \in S_3$ or $\mathbf{r}_e = \mathbf{1}_{++} \in D_4$ 
denotes the trivial representation, and $\mathbf{r}_a = \mathbf{1}' \in S_3$ or 
$\mathbf{r}_a = \mathbf{1}_{-+} \in D_4$ denotes the a non-trivial singlet representation.

 The scalar fields after electroweak symmetry breaking are given by
\begin{align}
    H_1=\frac{1}{\sqrt{2}}\begin{pmatrix} \sqrt{2}\phi^{+}_1 \\ v_{H_1}+R_{1}+i I_{1}  \end{pmatrix},
    H_2=\frac{1}{\sqrt{2}}\begin{pmatrix} \sqrt{2}\phi^{+}_2 \\ v_{H_2}+R_{2}+i I_{2}\end{pmatrix} \ \ \text{and} \ \
    \eta=\frac{1}{\sqrt{2}}\begin{pmatrix} \sqrt{2}\phi^{+}_3 \\ R_{3}+i I_{3}\end{pmatrix}
\end{align}

where the vevs $v_{H_1}$ and $v_{H_2}$ define the electroweak sphere,
\begin{align}
    v_{H_1}^2 + v_{H_2}^2 = v_{\rm EW}^2, \ \ \text{with}\ \ v_{\rm EW} = 246\,\text{GeV}.
\end{align}
In this analysis we got use of $\tan\beta = v_{H_2}/v_{H_1}$. The correspondence between field in the flavor basis and the mass basis is given in Table~\ref{tab:fmbasis}.

\begin{table}[h!]
\centering
\begin{tabular}{|c|c|c|}
\hline
 & \textbf{Flavor basis} & \textbf{Mass basis} \\
\hline

\textbf{Scalars} & $\underbrace{\{R_1,R_2,I_1,I_2,\phi^{\pm}_1,\phi^{\pm}_2\}}_{Z_2-\text{even}},\underbrace{\{R_3,I_3,\phi^{\pm}_3\}}_{Z_2-\text{odd}}$ & $\underbrace{\{h^0_1,h^0_2,G^A,h^A_2,G^{\pm},h^{\pm}_2\}}_{Z_2-\text{even}},\underbrace{\{\eta^0,\eta^A,\eta^{\pm}\}}_{Z_2-\text{odd}}$\\[20pt]
\hline
\end{tabular}
\caption{Scalar fields in the flavor and mass bases, where $G^A$ and $G^\pm$ are the neutral and charged Goldstone bosons, respectively.}
\label{tab:fmbasis}
\end{table}

For the active $Z_2$-even scalar fields, we define the mass eigenstates 
through the rotation matrices $O_R$, $O_I$, and $O_\pm$ as
\begin{align}
    \begin{pmatrix}
        h_1^0 \\ h_2^0
    \end{pmatrix} = O_R \begin{pmatrix}
        R_1 \\ R_2
    \end{pmatrix}, && 
    \begin{pmatrix}
       G^A \\ h^A_2
    \end{pmatrix} = O_I \begin{pmatrix}
       I_1 \\ I_2
    \end{pmatrix}, && 
    \begin{pmatrix}
        G^\pm \\ h^\pm_2
    \end{pmatrix} = O_\pm \begin{pmatrix}
     \phi_1^\pm \\  \phi_2^\pm   
    \end{pmatrix},
    \label{F-Pscalar}
\end{align}
satisfying
\begin{equation}
O_R\, M_R^2\, O_R^T = \text{diag}(m_{h^0_1}^2,\, m_{h^0_2}^2),~
O_I\, M_I^2\, O_I^T = \text{diag}(0,\, m_{h_2^A}^2),~
O_\pm\, M_\pm^2\, O_\pm^T = \text{diag}(0,\, m_{h_2^\pm}^2).
\end{equation}

\subsection{$S_3$ Scalar Potential}

\begin{align}
    V_{S_3} = V_0 + \lambda^{S_3}_7[(\Phi^\dagger\Phi)_{\mathbf{2}}(\Phi^\dagger\Phi)_{\mathbf{2}}]_{\mathbf{1}}
    + \frac{\lambda^{S_3}_8}{2}\left([(\Phi^\dagger H_1)_{\mathbf{2}}
    (\Phi^\dagger\Phi)_{\mathbf{2}}]_{\mathbf{1}} + h.c.\right).
\end{align}
The tadpole conditions give $\lambda_{456}=\lambda_4+\lambda_5+\lambda_6$, $c_\beta=\cos\beta$, $s_\beta=\sin\beta,$ and $t_\beta = \tan\beta$:
\begin{align}
    \mu^2_H &= v_{\rm EW}^2\left(c_\beta^2\lambda_1
    + \tfrac{1}{2}s_\beta^2\lambda_{456}
    - \tfrac{1}{4}s_\beta^2 t_\beta\lambda^{S_3}_8\right), \\
    \mu^2_\Phi &= v_{\rm EW}^2\left((\lambda_2+\lambda^{S_3}_7)s_\beta^2
    + \tfrac{1}{2}c_\beta^2\lambda_{456}
    - \tfrac{3}{4}s_\beta c_\beta\lambda^{S_3}_8\right).
\end{align}
Physical masses:
\begin{align}
    m^2_{h^0_1/h^0_2} &= v_{\rm EW}^2\!\left(2c_\beta^2\lambda_1
    + 2s_\beta^2(\lambda_2{+}\lambda^{S_3}_7)
    + \tfrac{1}{4}s_\beta(t_\beta s_\beta{-}3c_\beta)\lambda^{S_3}_8
    \mp\sqrt{\Delta_{S_3}}\right), \\
    m^2_{h^A_2} &= \tfrac{1}{4}v_{\rm EW}^2(t_\beta\lambda^{S_3}_8 - 4\lambda_6), \\
    m^2_{h^\pm_2} &= \tfrac{1}{4}v_{\rm EW}^2(t_\beta\lambda^{S_3}_8
    - 2(\lambda_5{+}\lambda_6)), \\
    m^2_{\eta^0} &= \tfrac{9}{4}v_{\rm EW}^2 s_\beta c_\beta\lambda^{S_3}_8, \\
    m^2_{\eta^A} &= \tfrac{1}{4}v_{\rm EW}^2(5s_\beta c_\beta\lambda^{S_3}_8
    - 8s_\beta^2(\lambda_3{+}\lambda^{S_3}_7) - 4c_\beta^2\lambda_6), \\
    m^2_{\eta^\pm} &= \tfrac{1}{4}v_{\rm EW}^2(5s_\beta c_\beta\lambda^{S_3}_8
    - 8s_\beta^2\lambda^{S_3}_7 - 2c_\beta^2(\lambda_5{+}\lambda_6)),
\end{align}
where $\Delta_{S_3} = (2c_\beta^2\lambda_1 - 2s_\beta^2(\lambda_2+\lambda^{S_3}_7)
+ s_\beta(t_\beta s_\beta+3c_\beta)\lambda^{S_3}_8)^2
+ 4(c_\beta s_\beta\lambda_{456} - \frac{3}{4}s_\beta^2\lambda^{S_3}_8)^2$.

\subsubsection*{Vacuum stability}
Copositivity conditions~\cite{Kannike:2012pe,motzkin1952}
(for $\lambda_5,\lambda_6<0$):
\begin{align}
    &\lambda_1>0,\quad \lambda_2+\lambda^{S_3}_7>0,\quad
    \lambda_4+\lambda_5-|\lambda_6|>-2\sqrt{\lambda_1(\lambda_2+\lambda^{S_3}_7)}, \\
    &\lambda_2>0 \;\text{ if }\; \lambda_3{+}\lambda^{S_3}_7>0,\lambda_3<0
    \;\text{ or }\; \lambda_3{+}\lambda^{S_3}_7<0,\lambda^{S_3}_7>0, \\
    &\lambda_2>\lambda_3 \;\text{ if }\; \lambda_3{+}\lambda^{S_3}_7>0,\lambda_3>0, \\
    &\lambda_1+\lambda_2+\lambda_3+\lambda_4+\lambda_5+\lambda_6+\lambda^{S_3}_7>\lambda^{S_3}_8.
\end{align}

\subsection{$D_4$ Scalar Potential}

\begin{align}
    V_{D_4} = V_0 + \lambda^{D_4}_7(\Phi^\dagger\Phi)_{+-}(\Phi^\dagger\Phi)_{+-}
    + \lambda^{D_4}_8(\Phi^\dagger\Phi)_{--}(\Phi^\dagger\Phi)_{--}.
\end{align}
The tadpole conditions give~\cite{Ivanov:2014doa,Bento:2022vsb}:
\begin{align}
    \mu^2_H &= \tfrac{1}{2}v_{\rm EW}^2
    \left(2c_\beta^2\lambda_1 + s_\beta^2\lambda_{456}\right), \\
    \mu^2_\Phi &= \tfrac{1}{2}v_{\rm EW}^2
    \left(2s_\beta^2(\lambda_2+\lambda^{D_4}_7) + c_\beta^2\lambda_{456}\right).
\end{align}
Physical masses:
\begin{align}
    m^2_{h^0_1/h^0_2} &= v_{\rm EW}^2\!\left(c_\beta^2\lambda_1
    + s_\beta^2(\lambda_2{+}\lambda^{D_4}_7)
    \mp\sqrt{\left(c_\beta^2\lambda_1{-}s_\beta^2(\lambda_2{+}\lambda^{D_4}_7)\right)^2
    + c_\beta^2 s_\beta^2\lambda_{456}^2}\right), \\
    m^2_{h^A_2} &= -v_{\rm EW}^2\lambda_6, \\
    m^2_{h^\pm_2} &= -\tfrac{1}{2}v_{\rm EW}^2(\lambda_5+\lambda_6), \\
    m^2_{\eta^0} &= 2s^2_\beta v_{\rm EW}^2(\lambda^{D_4}_8-\lambda^{D_4}_7), \\
    m^2_{\eta^A} &= -v_{\rm EW}^2(2s_\beta^2(\lambda_3+\lambda^{D_4}_7)
    + c_\beta^2\lambda_6), \\
    m^2_{\eta^\pm} &= -\tfrac{1}{2}v_{\rm EW}^2(4s_\beta^2\lambda^{D_4}_7
    + c_\beta^2(\lambda_5+\lambda_6)).
\end{align}

\subsubsection*{Vacuum stability}
\begin{align}
    &\lambda_1>0,\quad \lambda_2+\lambda^{D_4}_7>0,\quad
    \lambda_4+\lambda_5-|\lambda_6|>-2\sqrt{\lambda_1(\lambda_2+\lambda^{D_4}_7)}, \\
    &\lambda_2>0 \;\text{ if }\; \lambda_3{+}\lambda^{D_4}_8>0,\lambda_3<0
    \;\text{ or }\; \lambda_3{+}\lambda^{D_4}_8<0,\lambda^{D_4}_8>0, \\
    &\lambda_2>\lambda_3 \;\text{ if }\; \lambda_3{+}\lambda^{D_4}_8>0,\lambda_3>0, \\
    &\lambda_2>-\lambda^{D_4}_8 \;\text{ if }\; \lambda_3{+}\lambda^{D_4}_8<0,\lambda^{D_4}_8<0.
\end{align}

\section{One-Loop Mass Contributions}
\label{app:loop}

\subsection{Radiative Mass Contributions}

The mass generation at one-loop level proceeds through the two-vertex 
diagram with neutral spinor and scalar mediators~\cite{Escribano:2020iqq}. 
In our case, with no mixing among the $N$ fields, the contribution from 
the active scalar loops to the light neutrino mass matrix reads
\begin{align}
\bigl(m_\nu^{\text{active}}\bigr)_{\alpha\beta} = -\frac{1}{32\pi^2}
\sum_{f}\sum_{S\in\{h,A\}}\sum_{i,b,c}
\kappa_S^2\,(y_f)_{\alpha b}\,(y_f)_{\beta c}\,m_f\,
(O_S)_{bi}(O_S)_{ci}\,
B_0\!\left(0,\, m_{S_i}^2,\, m_f^2\right),
\label{eq:mnu-active}
\end{align}
and the dark scalar loops contribute

\begin{align}
\bigl(m_\nu^{\text{dark}}\bigr)_{\alpha\beta} = -\frac{m_{N_2}}{32\pi^2}
\sum_{D\in\{\eta^0,\eta^A\}}
\kappa_D^2\,y^D_{\alpha}\,y^D_{\beta}\,
B_0\!\left(0,\, m_D^2,\, m_{N_2}^2\right),
\label{eq:mnu-dark}
\end{align}
where $\alpha,\beta\in\{e,\mu,\tau\}$ are charged-lepton flavor indices, 
$f\in\{N_1,N_S\}$ runs over the neutral fermion mediators, 
$S\in\{h,A\}$ labels the active CP-even ($h$) and CP-odd ($A$) towers 
with mass eigenstates $h_i\in\{h_1^0,h_2^0\}$ and $A_i\in\{G^A,h_2^A\}$, 
$D\in\{\eta^0,\eta^A\}$ labels the dark CP-even and CP-odd scalars 
(which do not mix among themselves), and $b,c\in\{1,2\}$ are 
scalar-doublet flavor indices. The Yukawa couplings $(y_f)_{\alpha b}$ 
are written in the flavor basis. In our model
\begin{align}
    (y_{N_1})_{\alpha b} \equiv y^D_{\alpha}, 
    && 
    (y_{N_S})_{\alpha b} \equiv y^S_{\alpha}.
\end{align}
The matrices $(O_S)_{bi}$ are the entries of the rotation matrices 
defined in Eq.~\eqref{F-Pscalar} that connect the scalar flavor and 
mass bases, and the CP factor is $\kappa_S^2 = +1$ for CP-even and 
$\kappa_S^2 = -1$ for CP-odd scalars (likewise for $\kappa_D^2$).
\\

The Passarino--Veltman function $B_0$ at vanishing external 
momentum~\cite{Passarino:1978jh} is given, in dimensional 
regularization, by
\begin{align}
B_0(0, m_a^2, m_f^2) = \Delta_\epsilon + 1
- \frac{m_a^2 \ln\!\bigl(m_a^2/\mu^2\bigr) - m_f^2 \ln\!\bigl(m_f^2/\mu^2\bigr)}{m_a^2 - m_f^2},
\label{eq:B0}
\end{align}
where $\Delta_\epsilon = 2/\epsilon - \gamma_E + \ln 4\pi$ is the 
standard $\overline{\text{MS}}$ UV divergence and $\mu$ is the 
renormalization scale. Both $\Delta_\epsilon$ and the explicit $\mu$ 
dependence cancel exactly in each contribution to the neutrino mass 
matrix as a consequence of the orthogonality of the scalar rotation 
matrices,\footnote{The cancellation works as follows: the 
$\mu$-dependent piece of $B_0(0,m_a^2,m_f^2)$ is $-\ln\mu^2$, which 
upon summation factorizes the combination 
$\sum_i (O_S)_{bi}(O_S)_{ci} = \delta_{bc}$ for each fixed $S$. The 
same orthogonality argument applies to $\Delta_\epsilon$. The 
remaining sum over $S\in\{h,A\}$ with the relative sign $\kappa_S^2$ 
then ensures that the constant pieces (which would survive after the 
$\delta_{bc}$ projection) cancel between CP-even and CP-odd towers, 
leaving a finite, $\mu$-independent result. The same mechanism 
operates in the dark sector, where the rotation is trivial and the 
cancellation occurs directly between the $\eta^0$ and $\eta^A$ 
contributions. See Ref.~\cite{Escribano:2020iqq} for the general 
discussion in the multi-doublet scotogenic context.} ensuring that 
the neutrino mass matrix is finite and renormalization-scale 
independent.
\\

Grouping terms with identical Yukawa structure 
$(y_f)_{\alpha b}(y_f)_{\beta c}$, and using $m_{N_1}=m_{N_2}\equiv m_N$ 
and $m_{N_S}\equiv m_S$, we define the coefficients
\begin{align}
C_1 &= -\frac{m_N}{32\pi^2}\sum_{i=1}^{2}
\left[
(O_h)_{2i}^2\,B_0(0,m_{h_i}^2,m_N^2)
-
(O_A)_{2i}^2\,B_0(0,m_{A_i}^2,m_N^2)
\right],\\[4pt]
C_2 &= -\frac{m_N}{32\pi^2}
\left[
B_0(0,m_{\eta^0}^2,m_N^2)
-
B_0(0,m_{\eta^A}^2,m_N^2)
\right],\\[4pt]
C_3 &= -\frac{m_S}{32\pi^2}\sum_{i=1}^{2}
\left[
(O_h)_{1i}^2\,B_0(0,m_{h_i}^2,m_S^2)
-
(O_A)_{1i}^2\,B_0(0,m_{A_i}^2,m_S^2)
\right],
\end{align}
which enter the neutrino mass matrix.

\subsection{$Z$-Boson One-Loop Contributions}

The $Z$-boson loop contributes a term proportional to the type-I seesaw matrix \cite{AristizabalSierra:2011mn}:
\begin{align}
    M^Z_\nu = \frac{1}{m_f}\left(
    \frac{g^2}{64\pi^2 m_W^2}\left[
    m_{h^0_i}^2\ln\frac{m_f^2}{m_{h^0_i}^2}
    + 3m_Z^2\ln\frac{m_f^2}{m_Z^2}
    \right]\right) M_D^T M_D,
\end{align}
and therefore does not modify the neutrino mass spectrum or mixing angles.

\printbibliography

\end{document}